\documentclass[twocolumn,,showpacs,amssymb,amsmath,aps,pra]{revtex4-1}
\usepackage{amsmath}
\usepackage{amssymb}
\newcommand{\half}{\mbox{$\frac{1}{2}$}}
\usepackage{graphics}
\usepackage{bm}
\begin{document}
\title{Entanglement of finite cyclic chains at factorizing fields}
\author{R.Rossignoli, N. Canosa, J.M. Matera}
\affiliation{Departamento de F\'{\i}sica-IFLP,
Universidad Nacional de La Plata, C.C.67, La Plata (1900), Argentina}
\begin{abstract}
We examine the entanglement of cyclic spin 1/2 chains with anisotropic $XYZ$
Heisenberg couplings of arbitrary range at transverse factorizing magnetic
fields. At these fields the system exhibits a degenerate symmetry-breaking
separable ground state (GS). It is shown, however, that the side limits of the
GS pairwise entanglement at these fields are actually {\it non-zero} in finite
chains, corresponding such fields to a GS spin-parity transition. These limits
exhibit universal properties like being independent of the pair separation and
interaction range, and are directly related to the magnetization jump.
Illustrative exact results are shown for chains with I) full range and II)
nearest neighbor couplings. Global entanglement properties at such points are
also discussed.
\end{abstract}
\maketitle

Quantum entanglement is well recognized as a fundamental resource in quantum
information science \cite{NC.00}. It also provides a new perspective for the
analysis of quantum many-body systems, allowing to identify the genuine quantum
correlations \cite{ON.02,K.03,T.04}. An important result for quantum spin
chains with finite range interactions is that in contrast with the correlation
length, the {\it pairwise} entanglement range does not necessarily diverge at a
quantum phase transition \cite{ON.02}. For instance, it remains confined to
just first and second neighbors in a nearest neighbor Ising chain placed in a
transverse magnetic field \cite{ON.02}. It can, however, diverge at a different
point. Spin chains with anisotropic coupling exhibit a remarkable {\it
factorizing field} \cite{T.04,T.05,Ku.82}, where the system possesses a {\it
separable} GS and hence entanglement vanishes in principle, although it was
shown to reach {\it infinite range} in its vicinity \cite{Am.06}.

Previous analyses were focused on large systems. The purpose of this work is to
investigate the entanglement of {\it finite} chains, relevant for quantum
information processing, with anisotropic $XYZ$ coupling of {\it arbitrary}
range, exactly {\it at} transverse factorizing fields. Although it may seem
that any type of entanglement will vanish at such points, it should be noticed
that separable ground states break a fundamental symmetry of these chains (the
$S_z$ parity or global phase flip \cite{ON.02}) and are hence degenerate (and
non-orthogonal), the factorizing field corresponding actually to a GS
transition between opposite parity states.  As a consequence, pairwise
entanglement in finite chains will be shown to approach {\it distinct non-zero
side limits} at the factorizing field, which do not depend on the pair
separation or coupling range and whose average is directly measurable through
the magnitude of the magnetization jump at the transition. Moreover, even the
projector onto the GS subspace remains entangled at these fields. Although
these effects become negligible in large anisotropic chains, where opposite
parity ground states are nearly degenerate, they will be shown to be quite
prominent in small finite chains and, moreover, to remain appreciable for
increasing sizes if the $XY$ anisotropy becomes sufficiently small. Present
results are therefore particularly relevant for chains close to the $XXZ$
limit. We derive first general exact results valid for any range, describing
then illustrative exact results for chains with full range and nearest neighbor
couplings. The exact results in the last case are obtained through the
Jordan-Wigner mapping and its analytic parity dependent diagonalization.

We consider a cyclic chain of $n$ qubits or spins interacting through an
$XYZ$ Heisenberg coupling with arbitrary common range in a transverse magnetic
field $b$. Denoting with $\bm{s^i}$ the spin at site $i$, the Hamiltonian reads
\begin{eqnarray}
H&=&bS_z-\sum_{i<j}r_{j-i}(v_x s_x^is_x^{j}+v_y s_y^is_y^{j}+v_z s_z^is_z^{j})
\,,\label{H1}\\
&=&bS_z-\sum_{i<j}r_{j-i}[\half(v_+s_+^is_-^j+v_-s_+^is_+^j+h.c.) +v_z
s_z^is_z^{j}]\,,
 \nonumber\end{eqnarray}
where $\bm{S}=\sum_{i=1}^n \bm{s}^i$, $v_{\pm}=(v_x\pm v_y)/2$ and
$r_l=r_{n-l}$ for $l=j-i=1,\ldots,n-1$. Without loss of generality we can here
assume $b\geq 0$ and $v_x\geq |v_y|$ (i.e., $v_{\pm}\geq 0$), with $r_l$
arbitrary. We will be interested in the attractive (ferromagnetic) case
$r_l\geq 0$ $\forall$ $l$, although the following considerations are general.
Since $H$ conserves the $S_z$-parity,
\begin{equation}
[H,P_z]=0,\;\;P_z=\exp[i\pi (S_z+n/2)]\,,\label{P}
\end{equation}
its nondegenerate eigenstates will have definite parity $P_z=\pm 1$.

Let us now examine the conditions for which a
completely symmetric separable state of the form
\begin{equation}
|\theta\rangle=\prod_{i=1}^n
(\cos\half\theta|\!\!\downarrow_i\rangle+\sin\half\theta|\!\!\uparrow_i\rangle)
 =\exp[i\theta S_y]|0\rangle\,,
 \label{Phi}\end{equation}
where $s_z^i|\!\!\downarrow_i\rangle=-\half|\!\!\downarrow_i\rangle$ and
$|0\rangle=\prod_i|\!\!\downarrow_i\rangle$, can be an {\it exact} eigenstate
of (\ref{H1}). This state is fully aligned along an axes $z'$ forming an angle
$\theta$ with the $z$ axes, such that $S_{z'}|\theta\rangle=-\half
n|\theta\rangle$, {\it breaking parity symmetry} for $\theta\in(0,\pi)$.
Replacing $s_{z,x}^i=s_{z',x'}^i\cos\theta\pm s_{x',z'}^i \sin\theta$ in
(\ref{H1}), it is easily seen that these conditions are
\begin{eqnarray}
\cos\theta&=&\pm\sqrt{\chi},\;\;\;\chi\equiv\frac{v_y-v_z}{v_x-v_z},
\label{eq1}\\
b&=& r(v_x-v_z)\cos\theta,\;\;\;r\equiv \half\sum_{l=1}^{n-1}r_l\,,\label{eq2}
\end{eqnarray}
where Eq.\ (\ref{eq2}) is required for $\theta\in(0,\pi)$, i.e., $\chi\in[0,1)$
(in the $XXZ$ case $\chi=1$ ($v_y=v_x$) both $|0\rangle$ and $|\pi\rangle$ are
trivial eigenstates of (\ref{H1}) for {\it all} fields $b$). Such parity
breaking separable eigenstate is then feasible for $\chi\in [0,1)$ (i.e.,
$v_z\leq v_y<v_x$ if $v_x>|v_y|$) and $b=\pm b_s$, with
\begin{equation}
 b_s\equiv r(v_x-v_z)\sqrt{\chi}\,,\label{bs}
 \end{equation}
the {\it factorizing field}. The state (\ref{Phi}) will depend on the
anisotropy $\chi$ but not on the factors $r_l$, being then {\it independent} of
the interaction range.

It is also apparent that  {\it both} $|\theta\rangle$ and
$|\!-\theta\rangle=P_z|\theta\rangle$ are degenerate eigenstates of $H$ at
$b=b_s$, with energy
\begin{eqnarray}
\langle\theta| H|\theta\rangle&=&-\half n[b\cos\theta+
\half r(v_x\sin^2\theta+v_z\cos^2\theta)]\nonumber\\
 &=&-{\textstyle\frac{1}{4}} nr(v_x+v_y-v_z)\,.\label{enrg}\end{eqnarray}
Hence, at $b=\pm b_s$ two levels of opposite parity necessarily cross, enabling
the formation of these eigenstates.

Let us remark that in the attractive case $r_l\geq 0$ $\forall l$, the state
that minimizes $\langle H\rangle$ among separable states (i.e., the {\it mean
field} approximate GS) is precisely of the form (\ref{Phi}) $\forall\,b$, with
$|\theta|$ determined by Eq.\ (\ref{eq2}) if $|b|<b_c= r(v_x-v_z)$
(parity-breaking solution) and $\theta=0$ otherwise. Hence, in this case the
factorizing field can be seen as that where the mean field GS becomes an {\it
exact} eigenstate (i.e., the exact GS, as shown below).

The states $|\!\!\pm\theta\rangle$ will then
form a basis of the corresponding eigenspace at $b=b_s$ (assumed of dimension
$2$), which is {\it non-orthogonal} for $\theta\neq \pi/2$: $\langle
-\theta|\theta\rangle=\cos^n\theta$. A proper orthonormal basis conserving
parity symmetry is provided by the {\it entangled} states
 \begin{subequations}
\label{Psi}
\begin{eqnarray}
|\theta_\pm\rangle&\equiv&\frac{|\theta\rangle\pm |\!\!-\theta\rangle}
{\sqrt{2(1\pm \cos^n\theta)}}\label{Psia}\\
&=&\!\sum_{k {{\rm\;even}\atop {\rm\;odd}}}
\!\!\!\frac{\sqrt{2}\sin^k\!\frac{\theta}{2}\cos^{n-k}\!\frac{\theta}{2}}
{k!\sqrt{1\pm \cos^n\theta}}S_+^k|0\rangle,\label{Psib}
\end{eqnarray}
\end{subequations}
which satisfy $P_z|\theta_\pm\rangle=\pm|\theta_\pm\rangle$ {\it and are the
actual eigenstates of $H$ in each parity subspace at $b=b_s$}. These states
(and not the states $|\pm\theta\rangle$) are the actual limits of the
corresponding exact eigenstates $|\Psi^{\pm}(b)\rangle$ (which have definite
parity) for $b\rightarrow b_s$.

In the {\it attractive} case $r_{l}\geq 0$ $\forall$ $l$ (with $|v_y|\leq
v_x$), the states $|\theta_{\pm}\rangle$ (and hence $|\!\!\pm\theta\rangle$)
are {\it ground states} of $H$ at $b=b_s$:  The exact GS
$|\Psi_{0}^{\pm}(b)\rangle$ in each parity subspace must have expansion
coefficients all of the {\it same} sign in the standard computational basis
(i.e., that of separable states with definite values of $\{s_z^i\}$) in order
to minimize the average energy, since the average of the off-diagonal $XY$ term
in (\ref{H1}) can only increase (or eventually stay constant) for different
signs (as $r_{j-i}\geq 0$, $v_{\pm}\geq 0$) while those of the diagonal terms
$bS_z$ and $v_zs_z^is_z^j$ are sign independent. Hence,
$|\Psi_{0}^\pm(b)\rangle$ cannot be orthogonal to  $|\theta_{\pm}\rangle$,
whose expansion coefficients in this basis are all non-zero and of the same
sign  (Eq.\  (\ref{Psib})), and must then coincide with $|\theta_{\pm}\rangle$
at $b=b_s$.

Thus, in the attractive case $|\theta_{\pm}\rangle$ represent the {\it side
limits} $\lim_{b\rightarrow b_s^{\pm}}|\Psi_0(b)\rangle$ of the exact GS
$|\Psi_0(b)\rangle$ in the whole space at $b=b_s$, which undergoes there a
$|\theta_-\rangle\rightarrow |\theta_+\rangle$ parity transition (actually the
last parity transition as $b$ increases, as will be shown in the examples).

The pairwise entanglement in the states $|\theta_{\pm}\rangle$ depends
essentially on the {\it overlap} $\langle -\theta|\theta\rangle$. When
orthogonal ($\theta=\pi/2$), they are generalized GHZ states \cite{DVC.00},
which, although globally entangled, exhibit no pairwise entanglement (for
$n>2$). Moreover, in this case the normalized projector onto the space spanned
by the states $|\theta_{\pm}\rangle$,
\begin{equation}
\rho_0=\half(|\theta_+\rangle\langle\theta_+|+
|\theta_-\rangle\langle\theta_-|)\,,\label{rho0}
 \end{equation}
which represents in the attractive case the $T\rightarrow 0$ limit of the
thermal mixed state $\rho(T)\propto \exp[-H/kT]$ at $b=b_s$, is fully separable
(i.e., a convex combination of projectors onto separable states) as
$\rho_0=\frac{1}{2}(|\theta\rangle\langle\theta|
+|\!\!-\theta\rangle\langle-\theta|)$ for $\theta=\pi/2$. In contrast, for
$\theta\in(0,\pi/2)$ both states (\ref{Psi}) as well as the mixed state
(\ref{rho0}) will be shown to exhibit {\it a uniform non-zero entanglement
between any two spins} (note that the projector onto this subspace is no longer
the sum of the individual projectors $|\pm\theta\rangle\langle\pm\theta|$ when
$\langle-\theta|\theta\rangle\neq 0$).

Let us first evaluate the pairwise concurrence \cite{W.98} (a measure of
pairwise entanglement) in the states $|\theta_\pm\rangle$. As a consequence of
(\ref{P}) and the cyclic nature of $H$, the reduced two spin density matrix
$\rho_{ij}$ in any non-degenerate eigenstate or in $\rho(T)$, will commute with
the reduced parity $P^{ij}_{z}=e^{i\pi (s_z^i+s_z^j+1)}$ and depend just on
$l=|i-j|$. The ensuing concurrence $C_l\equiv C(\rho_{ij})$ takes the form
\begin{equation}
C_l=2\,{\rm Max}[|\alpha_l^+|-p_l,|\alpha_l^-|-q_l,0]\label{CC}\,,
 \end{equation}
where $\alpha_l^{\pm}=\langle s_+^is_{\pm}^j\rangle$, $p_l=\frac{1}{4}-\langle
s_z^is_z^j\rangle$, $q_l=[(\frac{1}{2}-p_l)^2-\langle s_z^i\rangle^2]^{1/2}$.
If $|\alpha_l^+|>p_l$ ($|\alpha_l^-|>q_l$) $C_l$ is of even (odd) parity type,
i.e., parallel (antiparallel) \cite{Am.06}, as in Bell state
$\propto|\!\!\uparrow\uparrow\rangle+|\!\!\downarrow\downarrow\rangle$
($|\!\!\uparrow\downarrow\rangle+|\!\!\downarrow\uparrow\rangle$). Just one of
these inequalities can be satisfied in a given state.

In the states (\ref{Psi}),
$\alpha_l^\nu=\frac{1}{4}\sin^2\theta\gamma^\nu_{\pm}$, $p_l=\alpha_l^-$,
$\langle s_z^i\rangle= -\half\cos\theta\gamma^+_{\pm}$, with
$\gamma^{\nu}_{\pm}= \frac{1\pm\nu\cos^{n-2}\theta}{1\pm\cos^n\theta}$ and
$\nu=\pm$.  We then obtain $C_{l}(|\theta_\pm\rangle)=C_{\pm}$ $\forall$ $l$,
with (assuming $\theta\in (0,\pi/2]$)
\begin{subequations}
\label{Cl}
 \begin{eqnarray}
 C_{\pm}&=&\sin^2\theta\frac{\cos^{n-2}\theta}{1\pm \cos^n\theta}\label{Cla}\\
&=&(1-\chi)\frac{\chi^{n/2-1}}{1\pm\chi^{n/2}}\,.\label{Clb}
\end{eqnarray}
\end{subequations}
Thus, $C_->C_+>0$, with $C_+$ ($C_-$) {\it parallel} ({\it antiparallel}). Note
that for $\theta\rightarrow 0$ ($\chi\rightarrow 1$),
\[C_+\rightarrow 0,\;\;\;C_-\rightarrow 2/n\,,\]
as in this limit $|\theta_+\rangle\rightarrow |0\rangle$ but
$|\theta_-\rangle\rightarrow|1\rangle\equiv \frac{1}{\sqrt{n}}S_+|0\rangle$,
which is an {\it $W$-state} \cite{DVC.00} ($2/n$ is in fact the {\it maximum}
value that can be attained by the concurrence in fully symmetric states
\cite{KBI.00}). As $\theta$ increases, $C_-$ decreases while $C_+$ becomes
maximum at $\theta\approx 1.6/\sqrt{n}$ (see Eq.\ \ref{r1}), vanishing both for
$\theta\rightarrow\pi/2$ ($\chi\rightarrow 0$) if $n>2$.

In the attractive case the values (\ref{Clb}) represent  the {\it universal
side limits} $C_{\pm}=\lim_{b\rightarrow b_s^{\pm}} C_l(b)$ of the GS
concurrences $C_l(b)$ at $b=b_s$, valid for {\it any} separation $l$ or
interaction range. For $\chi\rightarrow 1$ they correctly approach those for
the $|1\rangle\rightarrow |0\rangle$ transition taking place at $b=b_c$ in the
$XXZ$ limit \cite{CR.07} (where $b_s\rightarrow b_c$).

The concurrence jump $C_--C_+$ determines, noticeably, the concurrence
$C_0\equiv C_l(\rho_0)$ in the GS mixture (\ref{rho0}),
\begin{equation}
C_0=\half(C_--C_+)=(1-\chi)\frac{\chi^{n-1}}{1-\chi^n}\,,\label{C0}
\end{equation}
(see also Eq.\ \ref{ct}), which is of antiparallel type. It is a decreasing
function of $\theta$, starting at $1/n$ for $\theta\rightarrow 0$. In the
attractive case, Eq.\ (\ref{C0}) represents the common  $T\rightarrow 0$
limit of the thermal concurrences $C_l(T)$ at $b=b_s$ for any separation $l$
and coupling range.

\begin{figure}[t]
\vspace*{-0.cm}

\centerline{\scalebox{.55}{\includegraphics{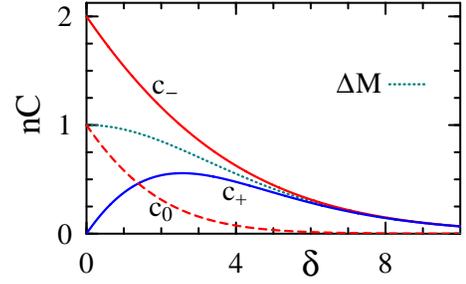}}}
 \vspace*{-.85cm}
\caption{(Color online) Ground state rescaled concurrences at the factorizing
field vs.\ scaled anisotropy parameter $\delta$. $c_{\pm}$ denote the side
limits (\ref{r1}), $c_0$ the value at $b=b_s$ (Eq.\ \ref{r2}), color indicating
the antiparallel ($c_-$, $c_0$) or parallel ($c_+$) type. The dotted line
depicts the magnetization jump (\ref{M}). These curves hold for any spin pair
and Hamiltonian of the form (\ref{H1}).} \label{f1}
\end{figure}

Although for fixed $\chi<1$, $C_{\pm}$ become exponentially small as $n$
increases, the rescaled concurrences $nC_{\pm}$ {\it remain finite for small
anisotropy} $\chi=1-\delta/n$. For large $n$ and fixed $\delta$, we obtain from
(\ref{Cl})--(\ref{C0}) the $n$-independent limits
\begin{eqnarray}
c_{\pm}&\equiv& nC_{\pm}\approx \delta e^{-\delta/2}/(1\pm e^{-\delta/2}),
\label{r1}\\
c_0&\equiv& n C_0\approx \delta e^{-\delta}/(1-e^{-\delta}),
 \label{r2}\end{eqnarray}
depicted in Fig.\ \ref{f1}. While $c_-$ and
$c_0=c_-(2\delta)/2$ are decreasing functions of $\delta$, $c_+$ is maximum at
$\delta=2[1+w(e^{-1})]\approx 2.56$, where $c_+=2w(e^{-1})\approx 0.56$ ($w(x)$
is the productlog function, such that $x=we^w$). We note also that $c_0>c_+$
for $\delta<2\ln 2$.

The mean rescaled concurrence $(c_++c_-)/2$ determines, remarkably, the {\it
total magnetization jump} at $b=b_s$:
\begin{subequations}
\label{M}
\begin{eqnarray}
\Delta M&\equiv&\langle\theta_-|S_z|\theta_-\rangle-\langle\theta_+
|S_z|\theta_+\rangle\nonumber\\
&=&n\sin^2\theta\frac{\cos^{n-1}\theta}{1-\cos^{2n}\theta}
=\half(c_++c_-)\sqrt{\chi},\label{dm}
\end{eqnarray}
which represents as well the slope of the {\it energy gap} $\Delta E\approx
(b-b_s)\Delta M$ between the odd and even GS at $b=b_s$. For large $n$ and
fixed $\delta$,
\begin{equation}\Delta M\approx (c_++c_-)/2=\delta
e^{-\delta/2}/(1-e^{-\delta})\,,\label{Mb}\end{equation}
\end{subequations}
 remaining finite and providing a
direct way to determine the average rescaled concurrence at $b_s$.

As illustration, we now show exact results for the concurrence in I) a fully
connected chain with constant $r_l$ \cite{DV.05} and II) a chain with nearest
neighbor coupling ($r_l=\delta_{l,1}+\delta_{l,n-1}$). In I) we set
$r_l=2/(n-1)$ $\forall$ $l$ such that $r=1$ in I and II (Eq.\ \ref{eq2}). The
factorizing field (\ref{bs}) and the energy (\ref{enrg}) are then the same in I
and II for fixed $v_{x,y,z}$. We will consider $v_x>0$ and  $v_z=0$  ($XY$
case).

In I, the GS can be obtained numerically by diagonalizing $H$ in the subspace
of maximum total spin states (to which it belongs) as $[H,S^2]=0$:
\[H_I=bS_z-\sum_{\mu=x,y}v_\mu(S_\mu^2-{\textstyle\frac{1}{4}}n)/(n-1)\,.\]
The fixed parity GS is then of the form $\sum_{k^{\rm even}_{\rm
(odd)}}w_kS_+^k|0\rangle$, leading to $l$ independent elements
$\alpha_l^+=\langle S_+^2\rangle/c_n$, $\alpha_l^-=(n^2/4-\langle
S_z^2\rangle)/c_n$, $\langle s_z^is_z^{i+l}\rangle=(\langle
S_z^2\rangle-n/4)/c_n$, with $c_n=n(n-1)$.

In II, the Hamiltonian can be solved {\it analytically} for any finite $n$ by
means of the Jordan-Wigner transformation \cite{LSM.61}, which allows to
rewrite $H$, {\it for each value ($\pm$) of the parity $P_z$}, as a quadratic
form in fermion operators $c^\dagger_i$, $c_i$ defined by
$c^\dagger_i=s_i^+\exp[-i\pi \sum_{j=1}^{i-1}s_j^+s_j^-]$:
\begin{eqnarray} H^{\pm}_{II}&=&
\sum_{i=1}^n b(c^\dagger_ic_i-\half)-\half\eta^{\pm}_i(v_+c^\dagger_i c_{i+1}
+v_-c^\dagger_i c^\dagger_{i+1}+h.c.)\nonumber\\
&=&\sum_{k\in K_{\pm}}\!\!\lambda_k (a^\dagger_k a_k
 -\half),\;\label{qd}\end{eqnarray}
where $n+1\equiv 1$, $\eta^-_i=1$, $\eta^+_i=1-2\delta_{in}$ and
 \[\lambda_k^2=(b-v_+\cos\omega_k)^2+v_-^2\sin^2\omega_k, \;\;\;
 \omega_k=2\pi k/n\,, \]
with $K_+=\{\half,\ldots,n-\half\}$, $K_-=\{0,\ldots,n-1\}$  (i.e., $k$
half-integer (integer) for positive (negative) parity). The diagonal form
(\ref{qd}) is obtained through a discrete {\it parity-dependent} Fourier
transform \cite{CR.07} $c^{\dagger}_j=\frac{e^{i\pi/4}}{\sqrt{n}} \sum_{k\in
K_{\pm}} e^{-i\omega_k j}c'^\dagger_k$, followed by a BCS transformation
$c'^\dagger_k=u_k a^\dagger_k+v_ka_{n-k}$, $c'_{n-k}=u_k a_{n-k}-v_k
a^\dagger_k$ to quasiparticle fermion operators $a_k$, $a^\dagger_k$, with
$u_k^2,v_k^2=\half[1\pm(b-v_+\cos\omega_k)/\lambda_k]$. For $b\geq 0$ we set
$\lambda_k\geq 0$ for $k\neq 0$ and $\lambda_0=v_+-b$, such that the
quasiparticle vacuum in $H_{II}^-$ is odd and the lowest energies for each
parity are $E_{II}^{\pm}=-\half\sum_{k\in k_{\pm}}\lambda_k$. At
$b=b_s=\sqrt{v_xv_y}$ (Eq.\ (\ref{bs})), $\lambda_k=v_+-b_s\cos\omega_k$ and
$E_{II}^+=E_{II}^-=-nv_+/2$, in full agreement with Eq.\ (\ref{enrg}).

The concurrences in the fixed parity GS can be obtained from the contractions
$f_l\equiv\langle c^\dagger_ic_j\rangle_\pm-\half \delta_{ij}$,
$g_l\equiv\langle c^\dagger_i c^\dagger_j\rangle_\pm$ and the use of Wick's
theorem \cite{LSM.61}, leading to $\langle s_z^i\rangle=f_0$, $\langle
s_z^is_z^j\rangle=f_0^2-f_l^2+g_l^2$ and
$\alpha_l^{\pm}={\textstyle\frac{1}{4}[{\rm det}(A^+_l) \mp{\rm det}(A^-_l)]}$,
with $(A_l^\pm)_{ij}=2(f_{i-j\pm 1}+g_{i-j\pm 1})$ $l\times l$ matrices.

\begin{figure}
\vspace*{-3.2cm}

\centerline{\hspace*{1.cm}\scalebox{1.}{\includegraphics{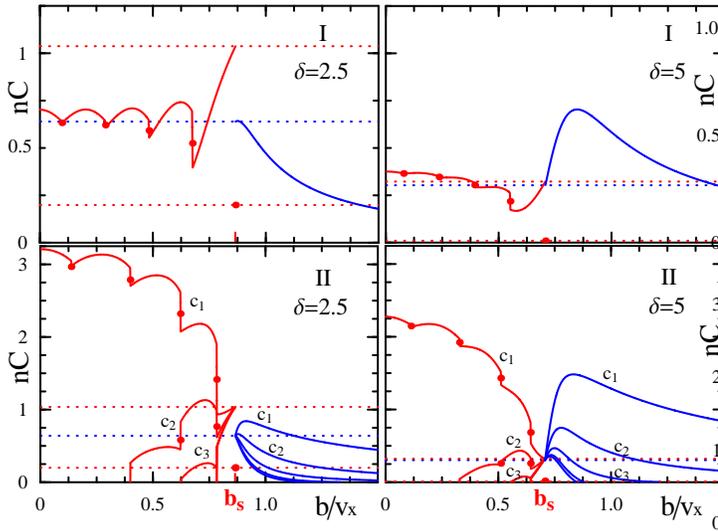}}}
\vspace*{-19.8cm}

\caption{(Color online) Ground state rescaled concurrences $c_l$ vs.\ magnetic
field $b$ for $n=10$ qubits and anisotropy $\chi=1-\delta/n=0.75$ (left) and
$0.5$ (right). Top panels correspond to a fully connected chain (case I), where
$c_l$ is the same for all separations $l$, bottom panels to nearest neighbor
coupling (case II), where $c_l$ is shown for all separations. Horizontal dotted
lines indicate the limit values (\ref{Cl}) and (\ref{C0}). At $b=b_s$,  $c_l$
changes from antiparallel (red) to parallel (blue) type, the side limits being
non-zero and identical $\forall$ $l$, and the same in I and II. Large dots
indicate concurrence values at the $n/2$ parity transitions (for the mixture of
both ground states), with that at $b=b_s$ given by Eq.\ (\ref{C0}) $\forall$
$l$ in I and II.} \label{f2}
\end{figure}

In both I and II, as $b$ increases from $0^+$, $[n/2]$ GS parity transitions
$\pm\rightarrow\mp$ take place if $\chi\in (0,1]$ (as in the $XXZ$ case
\cite{CR.07}), the last one $(-\rightarrow +$) at $b=b_s$. They are clearly
visible for low $n$ if $\chi$ is not small, i.e., if $\delta=n(1-\chi)$ is not
too large, as seen in Fig. \ref{f2} for $n=10$ qubits. For $b\rightarrow
b_s^\pm$, all GS concurrences $C_l^\pm$ are seen to approach the same side
limits (\ref{Cl}) in both I and II, which are non-negligible. For $\delta=2.5$,
$C_l^{\pm}$ reaches in fact its {\it maximum} for $b\rightarrow b_s\approx 0.87
v_x$ in I, and also in II if $l>2$. For $\delta=5$ the side limits are still
noticeable but nearly coincident, implying a negligible $c_0$ (Eq.\ \ref{C0}).

\begin{figure}[t]
\vspace*{-5.2cm}

\centerline{\hspace*{.7cm}\scalebox{1.}{\includegraphics{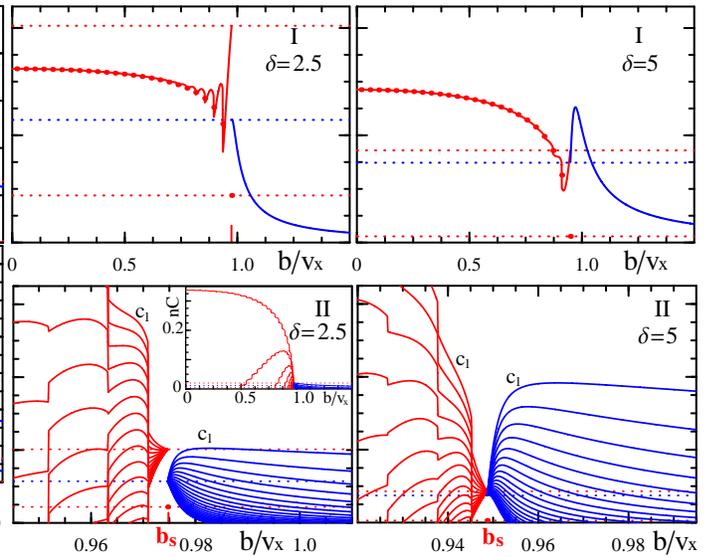}}}
 \vspace*{-17.5cm}

\caption{(Color online) Same details as Fig.\ \ref{f2} for $n=50$ qubits.
Bottom panels depict results in the vicinity of $b_s$, the inset those in the
full interval. The limit values at $b=b_s$ are again the same for all
separations and identical in I and II.}
 \label{f3}
\end{figure}

The behavior of $c_l^{\pm}$ for $n=50$ qubits at the same values of $\delta$
(now $\chi=0.95$ and $0.9$), depicted in Fig.\ \ref{f3}, is seen to be the same
as in Fig.\ \ref{f2} in the vicinity of $b_s$. All pairs become entangled as
$b\rightarrow b_s$, with $C_l$ approaching the common values (\ref{Cl}) in I
and II, now well approximated by Eqs.\ (\ref{r1}).

\begin{figure}[t]
\vspace*{-4.2cm}

\centerline{\hspace*{1.4cm}\scalebox{.8}{\includegraphics{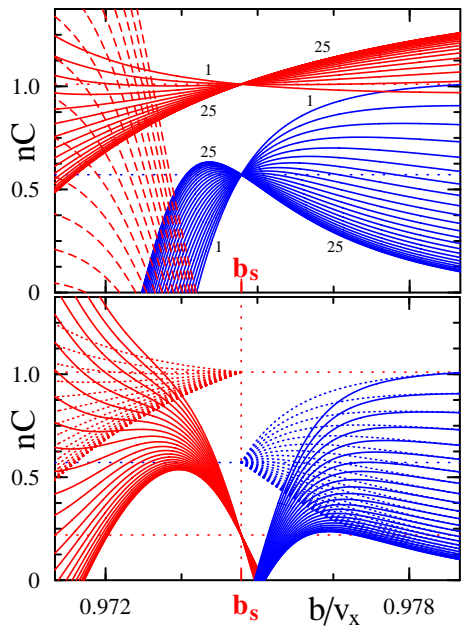}}}
 \vspace*{-11.5cm}

\caption{(Color online) Top: Rescaled concurrences $c_l$ in  the lowest state
for odd (upper curves, in red) and even (lower curves, in blue) parity $P$, for
case II with $\delta=2.5$ and $n=50$. Numbers indicate the separation $l$. At
$b=b_s$ an inversion in the ordering of the $c_l$'s with $l$ takes place.
Dashed lines depict the antiparallel concurrence in the even states. Bottom:
Thermal concurrences in the previous system at $kT=5\times 10^{-4} v_x$ (solid
lines) and at $T=0$ (dotted lines).}
 \label{f4}
\end{figure}

Within each parity subspace, the factorizing field is distinguished as that
where {\it all} GS concurrences $C_l^{\pm}$ {\it cross} at the values
(\ref{Cl}) (rather than vanish), as seen in the top panel of Fig.\ \ref{f4}.
Moreover, in II the {\it ordering} of concurrences $C_l^{\pm}$ becomes {\it
inverted}  at $b=b_s$: $C_l^-$ ($C_l^+$) {\it increases with increasing
separation $l$} for $b$ just above (below) $b_s$, as $C_l^{\pm}(b)$ is linear
close to $b_s$. Note  that $C_l^+(b)$ vanishes at a lower field $b_l^+<b_s$,
becoming {\it antiparallel} for $b<b_l^+$.

On the other hand, at sufficiently low temperatures, $b_s$ can be identified as
the field where all thermal concurrences $C_l(T)$ {\it cross} at the value
(\ref{C0}), as seen in the bottom panel of Fig.\ \ref{f4}. $C_l(T)$ vanishes at
a slightly {\it larger} $l$-dependent field $b_l(T)>b_s$, remaining {\it
antiparallel} until $b_l(T)$. To understand this effect, we note that in a
general mixture
\[\rho_q=q|\theta_+\rangle\langle\theta_+|
+(1-q)|\theta_-\rangle\langle\theta_-|,\;\;\;q\in[0,1]\]
the concurrence $C(q)\equiv C_l(\rho_q)$ is
\begin{equation}
 C(q)=|1-q/q_c|C_-,\;\;\;q_c=\half(1+\cos^n\theta)\,,
 \label{ct}\end{equation}
which generalizes Eq.\ (\ref{C0}) (recovered for $q=1/2$). $C(q)$ is
antiparallel (parallel) for $q<q_c$ ($>q_c$) {\it and zero at} $q=q_c>1/2$,
where $\rho_q$ becomes completely separable
($\rho_{q_c}=\half(|\theta\rangle\langle\theta|+
|\!\!-\theta\rangle\langle\!-\theta|)$). Separability requires then a slightly
{\it greater} weight in $|\theta_+\rangle$ due to its lower concurrence. Hence,
at low $T>0$ $C_l(T)$ vanishes and changes from antiparallel to parallel at a
slightly {\it larger} field $b_l(T)>b_s$, where the positive parity GS has a
higher weight in the thermal mixture. In case II this entails the surprising
result that in the narrow interval $b_s<b<b_1(T)$, the thermal concurrence
$C_l(T)$ {\it will increase with increasing $l$}, since it is still driven by
the lowest odd state $\forall$ $l$.

Let us finally mention that for any chain partition $(L,n-L)$, the Schmidt
number of the states (\ref{Psi}) is 2. Their Schmidt decomposition \cite{NC.00}
is
\begin{eqnarray}|\theta_\pm\rangle&=&
\sqrt{p_{L\pm}^+}|\theta_+^L\rangle|\theta_{\pm}^{n-L}\rangle+
\sqrt{p_{L\pm}^-}|\theta_-^L\rangle|\theta_\mp^{n-L}\rangle\,,\\
p_{L\pm}^\nu&=&\frac{(1+\nu\cos^L\theta)(1\pm\nu\cos^{n-L}\theta)}
 {2(1\pm\cos^n\theta)},\;\;(\nu=\pm) \label{pk}\end{eqnarray}
where $|\theta_{\pm}^L\rangle= (|\theta^L\rangle\pm
|\!\!-\theta^L\rangle)/\sqrt{2(1\pm \cos^L\theta)}$ denotes the analogous fixed
parity states for $L$ spins and $p_{L\pm}^\nu$ the eigenvalues of the ensuing
reduced density. The entanglement between $L$ and $n-L$ spins can be measured
through the entropy $S_L^\pm=-\sum_\nu p_{L\pm}^\nu\log_2 p_{L\pm}^\nu$ or
equivalently, the ``global'' concurrence $C_L^{\pm}=\sqrt{2(1-\sum_\nu
(p_L^\nu)^2)}$ \cite{RuC.03} (square root of the tangle for a pure state),
which is just an increasing function of $S_L^\pm$ ($S_L^\pm,C_L^\pm\in[0,1]$):
\[C_L^\pm= 2\sqrt{p_{L\pm}^+p_{L\pm}^-}=
 \frac{\sqrt{(1-\chi^L)(1-\chi^{n-L})}}{1\pm\chi^{n/2}} \,,\]
where we have replaced $\cos\theta=\sqrt{\chi}$. Hence, $C_L^-\geq C_L^+$, with
$C_L^\pm$ {\it increasing} functions of $\theta$, i.e., decreasing functions of
$\chi$, {\it in contrast} with the pairwise concurrences $C_{\pm}$. For
$\theta\rightarrow\pi/2$ ($\chi\rightarrow 0$), $C_L^\pm \rightarrow 1$ (GHZ
limit), whereas for $\theta\rightarrow 0$ ($\chi\rightarrow 1$),
$C_L^+\rightarrow 0$ but
\[C_L^-\rightarrow 2\sqrt{L(n-L)}/n\,,\]
($W$-state limit),  in which case  $S_L^-\approx (L/n)[1-\log_2(L/n)]$ for
$L\ll n$.  Thus, within the bounds imposed by a Schmidt number 2, the behavior
of $S_L^\pm$ and $C_L^\pm$ with $L$ is ``non-critical'' (i.e. saturated) for
low $\chi$ (large  $\delta$) and ``critical'' (non-saturated) for
$\chi\rightarrow 1$ (low $\delta$) and negative parity. It is also verified
that $C^\pm_1\geq \sqrt{n-1}C_{\pm}$ (in agreement with the general inequality
$C_{L=1}^2\geq \sum_{l=1}^{n-1}C_l^2$ \cite{CKW.00}), saturation reached for
$\theta\rightarrow 0$, where
$C^{\pm}_1/C_\pm\approx\sqrt{n-1}[1+\frac{1}{4}\theta^2(n-2)]$.

In summary, we have shown that due to the $S_z$ parity conservation, the GS of
finite cyclic chains with attractive couplings of the form (\ref{H1}) remains
entangled as the factorizing field $b_s$ is approached, undergoing at $b_s$ the
last parity transition and exhibiting for $b\rightarrow b_s^{\pm}$ universal
entanglement properties, ``intermediate'' between those of GHZ and W-states.
This field plays thus the role of a ``quantum critical field" for small chains,
with the pairwise concurrence reaching infinite range and approaching distinct
side limits which are independent of the pair separation and interaction range.
Their average is directly measurable through the GS magnetization jump
(\ref{dm}), which provides then a signature of the present effects, while their
difference determines the concurrence of the GS mixture (\ref{C0}). These
effects remain appreciable for increasing $n$ if the anisotropy becomes
sufficiently small (finite $\delta$), i.e., for chains close to the XXZ limit.
Moreover, within a fixed parity subspace (and also at sufficiently low $T>0$),
$b_s$ is singled out as the field where all pairwise concurrences cross, the
ordering with separation becoming inverted as $b$ crosses $b_s$. Type, range
and even ordering of the pairwise entanglement can thus be controlled by tuning
the field around $b_s$.

The authors acknowledge support from CIC (RR) and CONICET (NC,JMM) of
Argentina.

\end{document}